\documentstyle[12pt]{article}
\begin{document}
\begin{center}
{\large \bf Quark Mass Hierarchies and Maximal $CP$--Violation}
\end{center}
\bigskip
\begin{center}
{\bf Harald Fritzsch}
\end{center}
\begin{center}
{\bf Sektion Physik, Ludwig--Maximilians--Universit\"at}\\
{\bf D--80333 M\"unchen, Germany}
\end{center}

\bigskip
\bigskip
\noindent
{\bf Abstract}\\
{\small It is argued taking into account the observed mass spectra of the
leptons and
quarks that the phenomenon of flavor mixing is intimately related to the mass
spectra. We discuss a particularly interesting way to describe the flavor
mixing which is particularly suited for models of quark mass matrices based
on flavor symmetries. $CP$ violated is maximal or class to being maximal in
our approach.}\\

\noindent
At the end of the last century, exactly hundred years ago, physicists were
confronted with the puzzles of the atomic substructure and with the
perplexing structure of the atomic spectra. Finally, more than 20 years later,
all these complexities were understood in terms of a few parameters like
$\hbar , \alpha$ and the electron mass. Today we are facing similar
complexities -- the puzzles of the quark and lepton mass spectra and the
nature of flavor mixing. After the discovery of the $t$--quark the spectrum
of the masses (apart from the yet unknown neutrino
masses) is known. It is a rather wild spectrum, extending over 5 orders of
magnitude, from the tiny electron mass to the huge $t$--mass, but the actual
dynamics behind this spectrum remains mysterious. Nature speaks to us in some
kind of mysterious language. The letters of this language, i. e. the masses and
flavor mixing parameters, are known, but the grammar and the content of the
underlying text are unknown.\\
\\
Of course, in this lecture I cannot offer a
complete solution of the mass problem, but I shall describe what I would like
to define as the grammar of patterns and rules, which are not only very
simple, but seem to come well, if confronted with the experimental
results.\\
\\
The phenomenon of flavor mixing, which is intrinsically linked to
$CP$--violation, is an important ingredient of the Standard Model of Basic
Interactions. Yet unlike other features of the Standard Model, e.\ g.\ the
mixing of the neutral electroweak gauge bosons, it is a phenomenon which can
merely be described. A deeper understanding is still lacking, but it is
clearly directly linked to the mass spectrum of
the quarks. Furthermore there is a general consensus that a deeper dynamical
understanding would require to go beyond the physics of the Standard Model.
In my lectures I shall not go thus far. Instead I shall demonstrate that the
observed properties of the flavor mixing, combined with our knowledge about
the quark mass spectrum, suggest specific symmetry properties which allow
to fix the flavor mixing parameters with high precision, thus predicting the
outcome of the experiments which will soon be performed at the $B$--meson
factories.\\
\\
Before we enter the field of fermion mass generation, flavor mixing and
$CP$--violation, let me make some general remarks about the mass issue as it
appears today. The gauge interactions of the Standard Model are relevant both
for the lefthanded (L) and righthanded (R) fermion fields. Chirality is
conserved by the gauge interaction -- a lefthanded quark, after interacting
with a gauge boson, e. g. a $W$--boson or a gluon, stays lefthanded. A
$CP$--transformation turns a lefthanded quark into a righthanded antiquark,
but the interaction with the gauge bosons is unaffected. Thus the gauge
sector of the Standard Model can be divided into two disjoint worlds, the
world of $L$--fermions and of $R$--fermions. Formally the gauge interactions
do not provide a bridge between those two sectors.\\
\\
In reality the situation is more complex, which can be observed in particular
by looking at the strong interactions. In the limit in which the quarks are
taken to be massless (limit of chiral $SU(n)_L \times SU(n)_R$) the world of
QCD can also be divided up into the world of $L$--quarks and of $R$--quarks.
However nonperturbative effects generate a non--zero value for the v. e. v. of
$\bar q_R \, q_L$:
\begin{equation}
< 0 \mid \bar q_R \, q_L + h. c. \mid 0 > \not= 0,
\end{equation}
which is given in terms of $\Lambda_c$ ($\Lambda_c$: QCD scale). Thus there
exists a strong correlation betwen the lefthanded and
righthanded fields, which is responsible for the mass generation of the
bound states like the proton or the $\rho $--meson. These masses are due to
the dynamical breaking of the chiral symmetry.\\
\\
A consequence of this symmetry breaking is that the matrix elements of the
axial vector currents acquire a pole at $q^2 = 0$ ($q$: momentum
transfer), due to the massless pseudoscalar mesons which serve as the
corresponding Goldstone particles.\\
\\
In the Standard Model of the electroweak interaction the masses are
introduced by the coupling of the gauge fields and fermions to the scalar
field $\varphi $ whose neutral component $\varphi ^{\circ }$ acquires a
non--zero v.e.v.:
\begin{equation}
< 0 \mid \varphi ^{\circ } \mid 0 > = \frac{1}{\sqrt{2}} \, v
\end{equation}
In order to reproduce the observed gauge boson masses, one needs to have
$v \cong 246$ GeV.\\
\\
The quark and lepton masses are introduced by the coupling of the fermions
to $\varphi $, which is described by a coupling constant which is a free
parameter and varies for the different fermions in proportion to the masses.
These couplings of the type
\begin{equation}
\lambda \cdot \bar \psi_R \, \psi_L \, \varphi + {\rm h.c.}
\end{equation}
provide a correlation between the $L$--world and the $R$--world. The v.e.v. of
$\varphi$, multiplied with $\lambda $ describes the corresponding fermion
mass. Since the coupling constants $\lambda $ can be complex, the
$CP$--symmetry will be violated, if there are more than two families of
fermions, and if flavor mixing is present.\\
\\
In the Standard Model the fermion masses are introduced via the spontaneous
symmetry breaking in order to ensure the renormalizability of the underlying
gauge theory. However, it can be seen from a more general point of view that
the introduction of the fermion masses in the electroweak gauge theory is a
dynamical issue, unlike the introduction of the quark masses in QCD. Let us
consider a ``gedankenexperiment'', the process
$\bar t t \rightarrow W^+ W^-$, in which both incoming quarks are polarized.
In the center of mass frame we prepare the outgoing $W$--bosons in a
$J = 1$ wave by colliding both a $t_L$--quark and a $\bar t_R$--quark. The
tree-diagrams describing the process are either the formation of a virtual
$\gamma $ or $Z$, decaying into the $W$--pair, or the exchange of a
$b$--quark in the $t$--channel, leading to the production of the $W$--pair.
Both diagrams, if considered in isolation, lead to a cross section which
violates the unitarity bound for $J = 1$ at high energy, but the coherent sum
does not. This is the famous gauge theory cancellation.\\
\\
The dynamical aspect of the $t$--mass enters, if we study the
$W^+ W^-$--production in the $J = 0$ wave by considering the process
$t_L \, \bar t_L \rightarrow W^+ W^-$. Since $\bar t_L$--quarks do not
interact with the $W$--bosons, the cross section in the $J=0$ wave would
vanish for massless $t$--quarks. However, due to the non--zero $t$--mass, a
$t$--quark prepared in the
center--of--mass system with its spin opposite to its momentum has a
righthanded component, and the scattering amplitude in the $s$--wave is
proportional to $m_t \cdot \sqrt{s}$. Thus unitarity is violated at high
energy.\\
\\
In the Standard Model this problem is avoided, since there is a
cancellation in the $J = 0$ channel provided by the scalar
``Higgs''--particle.
The coupling of the latter to the $t$--quark is proportional to $m_t$.
Hence the cancellation is present, no matter how large $m_t$ is.\\
\\
This simple ``Gedankenexperiment'' shows the general condition: The cross
section for the reaction $\bar t t \rightarrow W^+ W$ in the $s$--wave must be
finite at high energies. This requires a new dynamics besides the one
provided by the quarks and electroweak gauge bosons. It could be either the
addition of a new scalar particle, as in the
Standard Model, or a string of resonances in the $J = 0$ channel, generated
by new
types of interactions or, perhaps, a new substructure of the leptons and
quarks. At present we do not know, which possibility is realized, but in
general it is implied that the lepton and quark masses are more than just
kinematical quantities. They must play an essential r$\hat{o}$le in the
dynamics. For this reason one should expect that the fermion masses,
especially the $t$--mass, are linked in a specific way to the masses of the
$W$--bosons, but the details of such a link are not known.\\
\\
After these introductory remarks about the r$\hat{o}$le of the lepton and
qark masses in the electroweak gauge theory, let me turn to the main topic
of these lectures, the connection between quark masses and the mixing of the
quark flavors. According to the standard electroweak theory one is dealing
with three $SU(2)_w$--doublets:
\begin{equation}
\left( \begin{array}{c}
	u'\\
	d' \end{array} \right)_L
\left( \begin{array}{c}
	c'\\
	s' \end{array} \right)_L
\left( \begin{array}{c}
	t'\\
	b' \end{array} \right)_L
\end{equation}
where $u', d' \ldots $ stand for certain superpositions of the corresponding
mass eigenstates. In terms of mass eigenstates the charged weak currents are
given by:
\begin{equation}
\overline{(u, c, t)_L} \left( \begin{array}{ccc}
				V_{ud} & V_{us} & V_{ub} \\
				V_{cd} & V_{cs} & V_{cb} \\
				V_{td} & V_{ts} & V_{tb}
			      \end{array} \right)
		     \left( \begin{array}{c}
			 d\\
			 s\\
			 s
			    \end{array} \right)_L \, .
\end{equation}
\\
This generalizes the standard Cabibbo--type rotation between the first and
second family$^{1)}$. The matrix elements
$V_{ij}$
are the elements of the CKM matrix$^{2)}$. In general they are complex
numbers. Their absolute
values are measurable quantities. For example, $|V_{cb}|$ primarily
determines the lifetime of $B$ mesons. The phases of $V_{ij}$,
however, are not physical, like the phases of quark fields. A phase
transformation of the $u$ quark ($u \rightarrow u ~ e^{{\rm
i}\alpha}$), for example, leaves the quark mass term invariant but
changes the elements in the first row of $V$ (i.e., $V_{uj} \rightarrow 
V_{uj} ~ e^{-{\rm i}\alpha}$). Only a common phase transformation of all 
quark fields leaves all elements of $V$ invariant, thus there is a
five-fold freedom to adjust the phases of $V_{ij}$.\\
\\
In general the unitary matrix $V$ depends on nine parameters.
Note that in the absence of complex phases $V$ would consist of only three 
independent parameters, corresponding to three (Euler) rotation
angles. Hence one can describe the complex matrix $V$ by three
angles and six phases. Due to the freedom in redefining the quark
field phases, five of the six phases in $V$ can be absorbed and we arrive
at the well-known result that the CKM matrix $V$ can be parametrized
in terms of three rotation angles and one $CP$-violating phase$^{2)}$.\\
\\
The standard parametrization of the CKM matrix is given as follows:
\begin{equation}
V_{ij} =
\left( \begin{array}{ccc}
	c_{12} \, c_{13} & s_{12} \, c_{13} & s_{13} \, e^{- i \delta _{13}}\\
	-s_{12} \, c_{23} - c_{12} \, s_{23} \, s_{13} e^{i \delta_{13}} &
	c_{12} \, c_{23} - s_{12} \, s_{23} \, s_{13} \, e^{i \delta_{13}}
	& s_{23} \, c_{13}  \\
	s_{12} \, s_{23} - c_{12} \, c_{23} \, s_{13} \, e^{i \delta_{13}} &
	-c_{12} \, s_{23} - s_{12} c_{23} s_{13} \, e^{i \delta_{13}} &
	c_{23} \, c_{13}
       \end{array} \right)
\end{equation}
\\
Here $s_{12}$ stands for $ {\rm sin} \Theta_{12}, c_{12}$ for ${\rm cos}
\Theta_{12}$ etc. Since the observed mixing angles are small the three angles
$\Theta_{12}, \Theta_{23}$ and $\Theta_{13}$ are related in a good
approximation to the moduli of specific $V$--elements as follows:
\begin{equation}
\mid V_{us} \mid \cong s_{12} \, , \;\;
\mid V_{ub} \mid \cong s_{13} \, , \;\;
\mid V_{cb} \mid \cong s_{23} \, .
\end{equation}
\\
The experiments give$^{3)}$:
\begin{equation}
\Theta_{12} \cong 12.7^{\circ } \; ,
\quad \Theta_{13} \cong 0.18^{\circ } \; ,
\quad \Theta_{23} \cong 2.25^{\circ} \; .
\end{equation}
(Here we have given the central values of these angles for illustration,
without indicating the errors. The phase $\delta_{13}$ angle will be
discussed later).\\
\\
Another way to describe the flavor mixing matrix is to follow
Wolfenstein$^{4)}$ and to use the modulus of $V_{us}$ as an expansion
parameter:
\begin{equation}
V = \left( \begin{array}{ccc}
	   1 - \frac{1}{2} \lambda^2 & \lambda &   A \lambda ^3 (\rho - i
	   \eta ) \\
	   - \lambda & 1 - \frac{1}{2} \lambda ^2 & A \, \lambda ^2 \\
	   A \, \lambda ^3 (1 - \rho - i \eta) & - A \, \lambda ^2 & 1
	   \end{array} \right) + O \left( \lambda ^4 \right)
\end{equation}
The central values of the parameters are:
\begin{equation}
\lambda = 0.2205 \; , \quad A = 0.806 \; , \quad \mid \rho - i \eta \mid = 0.36 \; .
\end{equation}
\\
When the standard parametrization of the CKM--matrix in terms of the angles
$\Theta_{ij}$ was introduced years ago by a number of authors including this
one$^{5)}$, the large value of the $t$--mass was not known. Thus the
striking mass hierarchy exhibited in the quark mass spectrum was not
explicitly taken into account. But the flavor mixing and the mass spectrum
are intimately related to each other, and the question arises whether the
standard way of describing the flavor mixing is the best way in doing so. We
shall discuss this issue below. The same question can be asked for the other
description proposed in the liberature, e. g. the original one given by
Kobayashi and Maskawa$^{2)}$ or the one given recently$^{6)}$.\\
\\
Adopting a particular parametrization
of flavor mixing is arbitrary and not directly a physical
issue. Nevertheless it is quite likely that the actual values of 
flavor mixing parameters (including the strength of $CP$--violation),
once they are known with high precision, will give interesting information 
about the physics beyond the standard model. Probably at this point it 
will turn out that a particular description of the CKM matrix is more
useful and transparent than the others. This is of special interest for the
phase parameter describing $CP$--violation. There are many different ways to
parametrize the phase parameter relevant for $CP$ violation, like there are
in principle many differenz ways to describe the parity violation in
nuclear transition amplitudes, for example by a phase angle. But only one
parametrization will be usefull at the end, the one which is intrinsically
linked to the physical origin of $CP$--violation. In the case of parity
violation it is the phase angle between the vector and axialvector currents
which is ninety degrees such that the weak interaction amplitude is given by
lefthanded quark fields. Later I shall show that something similar might be
true for $CP$--violation. Let me first 
analyze all possible parametrizations and point out their
respective advantages and disadvantages.\\
\\
The question about how many different ways to describe $V$ may exist was
raised some time ago$^{7)}$. Below we shall
reconsider this problem and give a complete analysis.\\
\\
If the flavor mixing matrix $V$ is first assumed to be a real orthogonal matrix, it can
in general be written as a product of three matrices $R_{12}$,
$R_{23}$ and $R_{31}$, which describe simple rotations in the (1,2),
(2,3) and (3,1) planes:
\begin{eqnarray}
R_{12}(\theta) & = & \left ( \matrix{
c^{~}_{\theta}  & s^{~}_{\theta}        & 0 \cr
- s^{~}_{\theta}        & c^{~}_{\theta}        & 0 \cr
0       & 0     & 1 \cr} \right ) \; , \nonumber \\ \nonumber \\
R_{23}(\sigma) & = & \left ( \matrix{
1       & 0     & 0 \cr
0       & c_{\sigma}    & s_{\sigma} \cr
0       & - s_{\sigma}  & c_{\sigma} \cr} \right ) \; , \nonumber \\
\nonumber \\
R_{31}(\tau) & = & \left ( \matrix{
c_{\tau}        & 0     & s_{\tau} \cr
0       & 1     & 0 \cr
- s_{\tau}      & 0     & c_{\tau} \cr} \right ) \; ,
\end{eqnarray}
where $s^{~}_{\theta} \equiv \sin \theta$, $c^{~}_{\theta} \equiv \cos
\theta$, etc.
Clearly these rotation matrices do not commute with each other.
There exist twelve different ways to arrange products of these
matrices such that the most general orthogonal matrix $R$ can be
obtained. 
Note that the matrix $R^{-1}_{ij} (\omega) $ plays an equivalent role
as $R_{ij} (\omega) $ in constructing $R$, because of $R^{-1}_{ij}(\omega) =
R_{ij}(-\omega)$. Note also that $R_{ij} (\omega) R_{ij}
(\omega^{\prime}) = R_{ij} (\omega + \omega^{\prime})$ holds, thus 
the product $R_{ij}(\omega) R_{ij}(\omega^{\prime})
R_{kl}(\omega^{\prime\prime})$ or $R_{kl}(\omega^{\prime\prime})
R_{ij}(\omega) R_{ij}(\omega^{\prime})$ cannot cover the whole space
of a $3\times 3$ orthogonal matrix and should be excluded.
Explicitly the twelve different forms of $R$ read as
\begin{eqnarray}
(1) & & R \; =\; R_{12}(\theta) ~ R_{23}(\sigma) ~ R_{12}(\theta^{\prime})
\; , \nonumber \\
(2) & & R \; =\; R_{12}(\theta) ~ R_{31}(\tau) ~ R_{12}(\theta^{\prime})
\; , \nonumber \\
(3) & & R \; =\; R_{23}(\sigma) ~ R_{12}(\theta) ~ R_{23}(\sigma^{\prime})
\; , \nonumber \\
(4) & & R \; =\; R_{23}(\sigma) ~ R_{31}(\tau) ~ R_{23}(\sigma^{\prime})
\; , \nonumber \\
(5) & & R \; =\; R_{31}(\tau) ~ R_{12}(\theta) ~ R_{31}(\tau^{\prime})
\; , \nonumber \\
(6) & & R \; =\; R_{31}(\tau) ~ R_{23}(\sigma) ~ R_{31}(\tau^{\prime})
\; , \nonumber
\end{eqnarray}
\newpage
\noindent
in which a rotation in the $(i,j)$ plane occurs twice;
and
\begin{eqnarray}
(7) & & R \; =\; R_{12}(\theta) ~ R_{23}(\sigma) ~ R_{31}(\tau)
\; , \nonumber \\
(8) & & R \; =\; R_{12}(\theta) ~ R_{31}(\tau) ~ R_{23}(\sigma)
\; , \nonumber \\
(9) & & R \; =\; R_{23}(\sigma) ~ R_{12}(\theta) ~ R_{31}(\tau)
\; , \nonumber \\
(10) & & R \; =\; R_{23}(\sigma) ~ R_{31}(\tau) ~ R_{12}(\theta)
\; , \nonumber \\
(11) & & R \; =\; R_{31}(\tau) ~ R_{12}(\theta) ~ R_{23}(\sigma)
\; , \nonumber \\
(12) & & R \; =\; R_{31}(\tau) ~ R_{23}(\sigma) ~ R_{12}(\theta) 
\; , \nonumber 
\end{eqnarray}
where all three $R_{ij}$ are present.\\ 
\\
Although all the above twelve combinations represent the most general 
orthogonal matrices, only nine of them are structurally different.
The rea-son is that the products $R_{ij} R_{kl} R_{ij}$ and $R_{ij} R_{mn} R_{ij}$ (with
$ij\neq kl\neq mn$) are correlated with each other, leading
essentially to the same form for $R$. Indeed it is straightforward to
see the correlation between patterns (1), (3), (5) and (2), (4),
(6), respectively, as follows:
\begin{eqnarray}
R_{12}(\theta) ~ R_{31}(\tau) ~ R_{12}(\theta^{\prime})
& = & R_{12}(\theta + \pi/2) ~ R_{23}(\sigma = \tau) ~
R_{12}(\theta^{\prime} - \pi/2) \; , \nonumber \\
R_{23}(\sigma) ~ R_{31}(\tau) ~ R_{23}(\sigma^{\prime})
& = & R_{23}(\sigma -\pi/2) ~ R_{12}(\theta = \tau) ~
R_{23}(\sigma^{\prime} + \pi/2) \; , \nonumber \\
R_{31}(\tau) ~ R_{23}(\sigma) ~ R_{31}(\tau^{\prime})
& = & R_{31}(\tau  + \pi/2) ~ R_{12}(\theta = \sigma) ~
R_{31}(\tau^{\prime} - \pi/2) \; .
\end{eqnarray}
Thus the orthogonal matrices (2), (4) and (6) need not be treated as
independent choices. 
We then draw the conclusion that
there exist {\it nine} different forms for the orthogonal matrix $R$,
i.e., patterns (1), (3) and (5) as well as (7) -- (12).\\ 
\\
We proceed to include the $CP$-violating phase, denoted by $\varphi$,
in the above rotation matrices. The resultant matrices should be
unitary such that a unitary flavor mixing matrix can be finally
produced. There are several different ways for
$\varphi$ to enter $R_{12}$, e.g., 
\[
R_{12} (\theta, \varphi) \; =\; \left ( \matrix{
c^{~}_{\theta}  & s^{~}_{\theta} ~ e^{+{\rm i} \varphi} & 0 \cr
- s^{~}_{\theta} ~ e^{-{\rm i} \varphi}         & c^{~}_{\theta}        & 0 \cr
0       & 0     & 1 \cr} \right ) \; , 
\]
or
\[
R_{12} (\theta, \varphi) \; =\; \left ( \matrix{
c^{~}_{\theta}  & s^{~}_{\theta}        & 0 \cr
- s^{~}_{\theta}        & c^{~}_{\theta}        & 0 \cr
0       & 0     & e^{-{\rm i} \varphi} \cr} \right ) \; , 
\]
or
\begin{equation}
R_{12} (\theta, \varphi) \; =\; \left ( \matrix{
c^{~}_{\theta} ~ e^{+{\rm i} \varphi}   & s^{~}_{\theta}        & 0 \cr
- s^{~}_{\theta}        & c^{~}_{\theta} ~ e^{-{\rm i} \varphi} & 0 \cr
0       & 0     & 1 \cr} \right ) \; . 
\end{equation}
Similarly one may introduce a phase parameter into $R_{23}$ or
$R_{31}$. Then the CKM matrix $V$ can be constructed, as a product of
three rotation matrices, by use of one complex $R_{ij}$ and two real ones. 
Note that the location of the $CP$-violating phase in $V$ can be arranged by
redefining the quark field phases, thus it does not play an essential role in 
classifying different parametrizations. We find that it is always
possible to locate the phase parameter $\varphi$ in a $2\times 2$ submatrix of
$V$, in which each element is a sum of two terms with the relative
phase $\varphi$. The remaining five elements of $V$ are real in such a 
phase assignment. Accordingly we arrive at nine distinctive
parametrizations of the CKM matrix $V$, where
the complex rotation matrices $R_{12}(\theta, \varphi)$,
$R_{23}(\sigma, \varphi)$ and $R_{31}(\tau, \varphi)$ are obtained
directly from the real ones in Eq. (11) with the replacement $1
\rightarrow e^{-{\rm i}\varphi}$. These nine possibilities have been
discussed recently in$^{8),9)}$.\\
\\
{\rm From} a mathematical point of view, all nine different parametrizations
are equivalent. However this is not the case if we apply our
considerations to the quarks and their mass spectrum. It is well--known that
both the observed
quark mass spectrum and the observed values of the flavor mixing
parameters exhibit a striking hierarchical structure. The latter can
be understood in a natural way as the consequence of a specific
pattern of chiral symmetries whose breaking causes the towers of
different masses to appear step by step$^{10), 11), 12)}$. Such a chiral
evolution of the mass matrices leads to a
specific way to introduce and describe the flavor mixing$^{11)}$.\\ 
\\
In the limit 
$m_u = m_d =0$, which is close to the real world, since $m_u/m_t \ll 1$ and
$m_d/m_b \ll 1$, the flavor mixing is merely a rotation between the
$t$--$c$ and $b$--$s$ systems, described by one rotation angle. No complex
phase is present; i.e., $CP$ violation is absent. This rotation angle
is expected to change very little, once $m_u$ and $m_d$
are introduced as tiny perturbations. A sensible parametrization should
make use of this feature. This implies that the rotation matrix
$R_{23}$ appears exactly once in the description of the CKM matrix
$V$, eliminating two among the nine possibilities. This leaves us with seven
parametrizations of the flavor mixing matrix.\\
\\
The list can be reduced further by considering the location of the phase $\varphi$. In
the limit $m_u = m_d =0$, the phase must disappear in the weak
transition elements $V_{tb}$, $V_{ts}$, $V_{cb}$ and $V_{cs}$. This
eliminates four possibilities. Then we are left with three
parametrizations. As expected, these are the
parametrizations containing the complex rotation matrix
$R_{23}(\sigma, \varphi)$. We stress that the ``standard'' parametrization
does not obey the above constraints and should be dismissed.\\
\\
Among the remaining three parametrizations, one parametrization is singled
out by 
the fact that the $CP$-violating phase $\varphi$ appears only in the
$2\times 2$ submatrix of $V$ describing the weak transitions among the 
light quarks. This is precisely the phase where the phase $\varphi$
should appear, not in any of the weak transition elements involving the 
heavy quarks $t$ and $b$.\\
\\
In the two other parametrizations the complex
phase $\varphi$ appears in $V_{cb}$ or $V_{ts}$, but this phase factor
is multiplied by a product of $\sin\theta$ and $\sin\tau$, i.e., it is of second 
order of the weak mixing angles. Hence the imaginary parts of these
elements are not exactly vanishing, but very small in magnitude.\\ 
\\
In our view the best possibility to describe the flavor mixing in the
standard model is to adopt the remaining parametrization$^{8)}$.
This parametrization has a number of significant
advantages in addition to that mentioned above. Especially it is well
suited for specific models of quark mass matrices.\\ 
\\
In the following part I shall show that this parametrization follows
automatically, if we impose the constraints from the chiral symmetries and the
hierarchical structure of the mass eigenvalues. We take the point of view
that the quark mass eigenvalues are dynamical entities, and one could change
their values in order to studay certain
symmetry limits, as it is done in QCD. In the standard electroweak model, in
which the quark mass matrices are given by the coupling of a scalar field to
various quark fields, this can certainly be done by adjusting the related
coupling constants. Whether it is possible in reality is an open question. It
is well--known that the quark mass matrices can always be made hermitian
by a suitable transformation of the right--handed fields. Without loss of
generality, we shall suppose in this paper that the quark mass matrices are
hermitian. In the limit where the masses of the $u$ and $d$ quarks are set to
zero, the quark mass matrix $\tilde{M}$ (for both charge $+2/3$ and
charge $-1/3$ sectors) can be arranged such that its elements 
$\tilde{M}_{i1}$ and $\tilde{M}_{1i}$ ($i=1,2,3$) are all zero$^{10), 11)}$.
Thus the quark mass matrices have the form
\begin{equation}
\tilde{M} \; =\; \left ( \matrix{
0       & 0     & 0 \cr
0       & \tilde{C}     & \tilde{B} \cr
0       & \tilde{B}^*   & \tilde{A} \cr} \right ) \; .
\end{equation}
The observed mass hierarchy is incorporated into this structure by
denoting the entry which is of the order of the $t$-quark or 
$b$-quark mass by $\tilde{A}$, with $\tilde{A}\gg \tilde{C},
|\tilde{B}|$. It can easily be seen$^{9)}$ that
the complex phases in the mass matrices (14) can be
rotated away by subjecting both $\tilde{M}_{\rm u}$ and
$\tilde{M}_{\rm d}$ to the
same unitary transformation. Thus we shall take $\tilde{B}$ to be
real for both up- and down-quark sectors. As expected, $CP$ violation
cannot arise at this stage. The diagonalization of the mass matrices
leads to a mixing between the second and third families, described by an
angle $\tilde{\theta}$. The flavor mixing matrix is
then given by
\begin{equation}
\tilde{V} \; =\; \left ( \matrix{
1       & 0     & 0 \cr
0       & \tilde{c}     & \tilde{s} \cr
0       & -\tilde{s}    & \tilde{c} \cr } \right ) \; ,
\end{equation}
where $\tilde{s} \equiv \sin \tilde{\theta}$ and $\tilde{c} \equiv
\cos \tilde{\theta}$. In view of the fact that the limit $m_u = m_d
=0$ is not far from reality, the angle $\tilde{\theta}$ is essentially 
given$^{13)}$ by the observed value of $|V_{cb}|$ ($=0.039 \pm 0.002$;
i.e., $\tilde{\theta} = 2.24^{\circ} \pm 0.12^{\circ}$.\\ 
\\
At the next and final stage of the chiral evolution of the mass matrices,
the masses of the $u$ and $d$ quarks are introduced.
The Hermitian mass matrices have in general the
form:
\begin{equation}
M \; =\; \left ( \matrix{
E       & D     & F \cr
D^*     & C     & B \cr
F^*     & B^*   & A \cr } \right ) \; 
\end{equation}
with $A\gg C, |B| \gg E, |D|, |F|$. By a common unitary transformation of 
the up- and down-type quark fields, one can always arrange the mass
matrices $M_{\rm u}$ and $M_{\rm d}$ in such a way that $F_{\rm u} =
F_{\rm d} =0$; i.e.,
\begin{equation}
M \; =\; \left ( \matrix{
E       & D     & 0 \cr
D^*     & C     & B \cr
0       & B^*   & A \cr } \right ) \; .
\end{equation}
This can easily be seen as follows. If phases are neglected, the two
symmetric mass matrices $M_{\rm u}$ and $M_{\rm d}$ can be transformed 
by an orthogonal transformation matrix $O$, which can be described by
three angles such that they assume the form (17). The condition
$F_{\rm u} =F_{\rm d} =0$ gives two constraints for the three angles of 
$O$. If complex phases are allowed in $M_{\rm u}$ and $M_{\rm d}$, the 
condition $F_{\rm u} =F_{\rm u}^* = F_{\rm d} =F_{\rm d}^* =0$ imposes
four constraints, which can also be fulfilled, if $M_{\rm u}$ and
$M_{\rm d}$ are subjected to a common unitary transformation matrix $U$. The
latter depends on nine parameters. Three of them are not suitable for
our purpose, since they are just diagonal phases; but the remaining
six can be chosen such that the vanishing of $F_{\rm u}$ and $F_{\rm
d}$ results.\\ 
\\
The basis in which the mass matrices take the form (17) is a basis in
the space of quark flavors, which in our view is of special
interest. It is a basis in which the mass matrices exhibit two
texture zeros, for both up- and down-type quark sectors. 
These, however, do not imply special relations among
mass eigenvalues and flavor mixing parameters (as pointed out
above). In this basis the mixing is of the ``nearest neighbour'' form, 
since the (1,3) and (3,1) elements of $M_{\rm u}$ and $M_{\rm d}$
vanish; no direct mixing between the heavy $t$ (or $b$) quark and the
light $u$ (or $d$) quark is present.
In certain models$^{14), 15)}$,
this basis is indeed of particular interest, but we shall proceed without 
relying on a special texture models for the mass matrices.\\
\\
A mass matrix of the type (17) can in the absence of complex phases be
diagonalized by a rotation matrix, described only by two angles in
the hierarchy limit of quark masses$^{8)}$.
At first the off-diagonal element 
$B$ is rotated away by a rotation between the second and third 
families (angle $\theta_{23}$); at the second step the element $D$ is rotated away by a
transformation of the first and second families (angle $\theta_{12}$). No rotation between
the first and third families is required to an excellent degree of 
accuracy. The rotation matrix for this sequence takes the form
\begin{eqnarray}
R \; =\; R_{12} R_{23} & = & \left ( \matrix{
c_{12}  & s_{12}        & 0 \cr
-s_{12} & c_{12}        & 0 \cr
0       & 0     & 1 \cr } \right )  \left ( \matrix{
1       & 0     & 0 \cr
0       & c_{23}        & s_{23} \cr
0       & -s_{23}       & c_{23} \cr } \right ) \; 
\nonumber \\ \nonumber \\
& = & \left ( \matrix{
c_{12}  & s_{12} c_{23} & s_{12} s_{23} \cr 
-s_{12} & c_{12} c_{23} & c_{12} s_{23} \cr
0       & -s_{23}       & c_{23} \cr } \right ) \; ,
\end{eqnarray}
where $c_{12} \equiv \cos \theta_{12}$, $s_{12} \equiv \sin
\theta_{12}$, etc.
The flavor mixing matrix $V$ is the product of two such matrices, one
describing the rotation among the up-type quarks, and the other describing
the rotation among the down-type quarks:
\begin{equation}
V \; =\; R^{\rm u}_{12} R^{\rm u}_{23} ( R^{\rm d}_{23} )^{-1} ( R^{\rm d}_{12} )^{-1} \; .
\end{equation}
Note that $V$ itself is exact, since a rotation between the first and
third families can always be incorporated and absorbed through redefining
the relevant rotation matrices.
The product $R^{\rm u}_{23} (R^{\rm d}_{23} )^{-1}$ can be written as
a rotation matrix described by a single angle $\theta$. In the limit
$m_u = m_d =0$, this is just the angle $\tilde{\theta}$ encountered
in Eq. (15). The angle which describes the $R^{\rm u}_{12}$ rotation shall
be denoted by $\theta_{\rm u}$; the corresponding angle for the
$R^{\rm d}_{12}$ rotation by $\theta_{\rm d}$. Thus in the absence of
$CP$-violating phases the flavor mixing matrix takes the following 
specific form:
\begin{eqnarray}
V & = & \left ( \matrix{
c_{\rm u}       & s_{\rm u}     & 0 \cr
-s_{\rm u}      & c_{\rm u}     & 0 \cr
0       & 0     & 1 \cr } \right )  \left ( \matrix{
1       & 0     & 0 \cr
0       & c     & s \cr
0       & -s    & c \cr } \right )  \left ( \matrix{
c_{\rm d}       & -s_{\rm d}    & 0 \cr
s_{\rm d}       & c_{\rm d}     & 0 \cr
0       & 0     & 1 \cr } \right )  \nonumber \\ \nonumber \\
& = & \left ( \matrix{
s_{\rm u} s_{\rm d} c + c_{\rm u} c_{\rm d}     & 
s_{\rm u} c_{\rm d} c - c_{\rm u} s_{\rm d}     & s_{\rm u} s \cr
c_{\rm u} s_{\rm d} c - s_{\rm u} c_{\rm d}     & 
c_{\rm u} c_{\rm d} c + s_{\rm u} s_{\rm d}     & c_{\rm u} s \cr
-s_{\rm d} s    & -c_{\rm d} s  & c \cr } \right ) \; ,
\end{eqnarray}
where $c_{\rm u} \equiv \cos\theta_{\rm u}$, $s_{\rm u} \equiv
\sin\theta_{\rm u}$, etc.\\
\\
We proceed by including the phase parameters of the quark mass
matrices in Eq. (17). Each mass matrix has in general two complex 
phases. But it can easily be seen that, 
by suitable rephasing of the quark fields,
the flavor mixing matrix can finally be written in terms of only a
single phase $\varphi$ as follows$^{6)}$:
\begin{eqnarray}
V & = & \left ( \matrix{
c_{\rm u}       & s_{\rm u}     & 0 \cr
-s_{\rm u}      & c_{\rm u}     & 0 \cr
0       & 0     & 1 \cr } \right )  \left ( \matrix{
e^{-{\rm i}\varphi}     & 0     & 0 \cr
0       & c     & s \cr
0       & -s    & c \cr } \right )  \left ( \matrix{
c_{\rm d}       & -s_{\rm d}    & 0 \cr
s_{\rm d}       & c_{\rm d}     & 0 \cr
0       & 0     & 1 \cr } \right )  \nonumber \\ \nonumber \\
& = & \left ( \matrix{
s_{\rm u} s_{\rm d} c + c_{\rm u} c_{\rm d} e^{-{\rm i}\varphi} &
s_{\rm u} c_{\rm d} c - c_{\rm u} s_{\rm d} e^{-{\rm i}\varphi} &
s_{\rm u} s \cr
c_{\rm u} s_{\rm d} c - s_{\rm u} c_{\rm d} e^{-{\rm i}\varphi} &
c_{\rm u} c_{\rm d} c + s_{\rm u} s_{\rm d} e^{-{\rm i}\varphi}   &
c_{\rm u} s \cr
- s_{\rm d} s   & - c_{\rm d} s & c \cr } \right ) \; .
\end{eqnarray}
Note that the three angles $\theta_{\rm u}$, $\theta_{\rm d}$ and
$\theta$ in Eq. (21) can all be arranged to lie in the first quadrant
through a suitable redefinition of quark field phases. Consequently
all $s_{\rm u}$, $s_{\rm d}$, $s$ and $c_{\rm u}$, $c_{\rm d}$, $c$
are positive. The phase $\varphi$ can in general take values from 0
to $2\pi$; and $CP$ violation is present in weak interactions
if $\varphi \neq 0, \pi$ and $2\pi$. The parametrization given above is the
one which is left after passing through the various steps of
elimination.\\
\\
This representation of the flavor mixing matrix,
in comparison with all other parametrizations discussed
previously, has a number of interesting features which in our view make it very
attractive and provide strong arguments for its use in future
discussions of flavor mixing phenomena, in particular, those in
$B$-meson physics. We shall discuss them below.\\
\\
a) The flavor mixing matrix $V$ in Eq. (21) follows directly from the
chiral expansion of the mass
matrices. Thus it naturally takes into account the hierarchical structure of the 
quark mass spectrum.

b) The complex phase describing $CP$ violation ($\varphi$) appears only in the
(1,1), (1,2), (2,1) and (2,2) elements of $V$, i.e., 
in the elements involving only the quarks of the first and second
families. This is a natural description of $CP$ violation since in our 
hierarchical approach $CP$ violation is not directly linked to the third family, but
rather to the first and second ones, and in particular to the mass terms of the
$u$ and $d$ quarks. 

It is instructive to consider the special case $s_{\rm u} = s_{\rm d}
= s = 0$. Then the flavor mixing matrix $V$ takes the form
\begin{equation}
V \; = \; \left ( \matrix{
e^{-{\rm i}\varphi}     & 0     & 0 \cr
0       & 1     & 0 \cr
0       & 0     & 1 \cr} \right ) \; .
\end{equation}
This matrix describes a phase change in the weak transition between
$u$ and $d$, while no phase change is present in the
transitions between $c$ and $s$ as well as $t$ and $b$.
Of course, this effect can be absorbed in a phase change of the $u$-
and $d$-quark fields, and no $CP$ violation is present. Once the
angles $\theta_{\rm u}$, $\theta_{\rm d}$ and $\theta$ are introduced, 
however, $CP$ violation arises. It is due to a phase change in the weak
transition between $u^{\prime}$ and $d^{\prime}$, where $u^{\prime}$
and $d^{\prime}$ are the rotated quark fields, obtained by applying
the corresponding rotation matrices given in Eq. (21) to the 
quark mass eigenstates ($u^{\prime}$: mainly $u$, small admixture of
$c$; $d^{\prime}$: mainly $d$, small admixture of $s$).

Since the mixing matrix elements involving $t$ or $b$ quark are real
in the representation (21), one can find that the phase parameter of
$B^0_q$-$\bar{B}^0_q$ mixing ($q=d$ or $s$), dominated by the
box-diagram contributions in the standard model$^{3)}$, is essentially
unity:
\begin{equation}
\left ( \frac{q}{p} \right )_{B_q} = \;
\frac{V^*_{tb}V_{tq}}{V_{tb}V^*_{tq}} \; = \; 1 \; .
\end{equation}
In most of other parametrizations of the flavor mixing matrix,
however, the two rephasing-variant quantities 
$(q/p)^{~}_{B_d}$ and $(q/p)^{~}_{B_s}$ take different (maybe complex) values.

c) The dynamics of flavor mixing can easily be interpreted by
considering certain limiting cases in Eq. (21). In the limit $\theta
\rightarrow 0$ (i.e., $s \rightarrow 0$ and $c\rightarrow 1$), the
flavor mixing is, of course, just a mixing between the first and
second families, described by only one mixing angle (the Cabibbo angle 
$\theta_{\rm C}$).  
It is a special and essential feature of the representation (21) that the Cabibbo
angle is {\it not} a basic angle, used in the parametrization. 
The matrix element $V_{us}$ (or $V_{cd}$) is
indeed a superposition of two terms including a phase. This feature
arises naturally in our hierarchical approach, but it is not new. In
many models of specific textures of mass matrices, it is indeed the
case that the Cabibbo-type transition $V_{us}$ (or $V_{cd}$) 
is a superposition of several
terms. At first, it was obtained by in the discussion of the two-family
mixing$^{16)}$.

In the limit $\theta =0$ considered here, one has $|V_{us}| = |V_{cd}|
= \sin\theta_{\rm C} \equiv s^{~}_{\rm C}$ and
\begin{equation}
s^{~}_{\rm C} \; =\; \left | s_{\rm u} c_{\rm d} ~ - ~ c_{\rm u} s_{\rm d}
e^{-{\rm i}\varphi} \right | \; .
\end{equation}
This relation describes a triangle in the complex plane, as
illustrated in Fig. 1, which we shall denote as the ``LQ-- triangle''
(``light quark triangle''). 
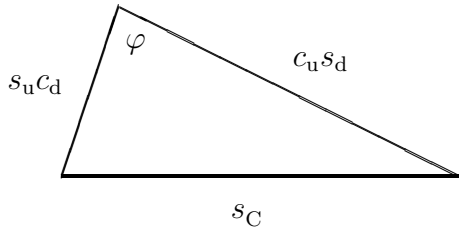
\begin{figure}[t]
\begin{picture}(400,160)(-20,210)
\put(80,300){\line(1,0){150}}
\put(80,300.5){\line(1,0){150}}
\put(150,285.5){\makebox(0,0){$s^{~}_{\rm C}$}}
\put(80,300){\line(1,3){21.5}}
\put(80,300.5){\line(1,3){21.5}}
\put(80,299.5){\line(1,3){21.5}}
\put(70,335){\makebox(0,0){$s_{\rm u} c_{\rm d}$}}
\put(230,300){\line(-2,1){128}}
\put(230,300.5){\line(-2,1){128}}
\put(178,343.5){\makebox(0,0){$c_{\rm u} s_{\rm d}$}}

\put(108,350){\makebox(0,0){$\varphi$}}
\end{picture}
\vspace{-2.5cm}
\caption{The LQ--triangle in the complex plane.}
\end{figure}
This triangle is a feature of
the mixing of the first two families. Explicitly one has (for $s=0$):
\begin{equation}
\tan\theta_{\rm C} \; =\; \sqrt{\frac{\tan^2\theta_{\rm u} +
\tan^2\theta_{\rm d} - 2 \tan\theta_{\rm u} \tan\theta_{\rm d}
\cos\varphi}
{1 + \tan^2\theta_{\rm u} \tan^2\theta_{\rm d} + 2 \tan\theta_{\rm u}
\tan\theta_{\rm d} \cos\varphi}} \; .
\end{equation}
Certainly the flavor mixing matrix $V$ cannot accommodate $CP$ violation in this
limit. However, the existence of $\varphi$ seems necessary in order
to make Eq. (25) compatible with current data, as one can see below.

d) The three mixing angles $\theta$, $\theta_{\rm u}$ and 
$\theta_{\rm d}$ have a precise physical meaning. The angle $\theta$
describes the mixing between the second and third families, which is
generated by the off-diagonal terms $B_{\rm u}$ and $B_{\rm d}$ in the 
up and down mass matrices of Eq. (17). 
We shall refer to this mixing involving $t$ and $b$ as the ``heavy
quark mixing''.
The angle $\theta_{\rm u}$,
however, solely describes the $u$-$c$ mixing, corresponding to the $D_{\rm
u}$ term in $M_{\rm u}$. We shall denote this as the ``u-channel mixing''.
The angle $\theta_{\rm d}$ solely describes 
the $d$-$s$ mixing, corresponding to the $D_{\rm d}$ term in $M_{\rm
d}$; this will be denoted as the ``d-channel mixing''. 
Thus there exists an asymmetry between the mixing of the first and
second families and that of the second and third families,
which in our view reflects interesting details of the underlying dynamics of
flavor mixing. 
The heavy quark mixing is a combined effect, involving both charge
$+2/3$ and charge $-1/3$ quarks, while the u- or d-channel mixing
(described by the angle $\theta_{\rm u}$ or $\theta_{\rm d}$) proceeds 
solely in the charge $+2/3$ or charge $-1/3$ sector. Therefore an
experimental determination of these two angles would allow to draw
interesting conclusions about the amount and perhaps the underlying
pattern of the u- or d-channel mixing.

e) The three angles $\theta$, $\theta_{\rm u}$ and $\theta_{\rm d}$
are related in a very simple way to observable quantities of $B$-meson 
physics. 
For example, $\theta$ is related to 
the rate of the semileptonic decay $B\rightarrow D^*l\nu^{~}_l$; 
$\theta_{\rm u}$ is associated with the ratio of the decay rate of
$B\rightarrow (\pi, \rho) l \nu^{~}_l$ to that of $B\rightarrow 
D^* l\nu^{~}_l$; and $\theta_{\rm d}$ can be determined from the ratio of
the mass difference between two $B_d$ mass eigenstates to that between
two $B_s$ mass eigenstates. We find the following exact
relations:
\begin{equation}
\sin \theta \; = \; |V_{cb}| \sqrt{ 1 + \left |\frac{V_{ub}}{V_{cb}}
\right |^2} \; ,
\end{equation}
and
\begin{eqnarray}
\tan\theta_{\rm u} & = & \left | \frac{V_{ub}}{V_{cb}} \right | \; ,
\nonumber \\
\tan\theta_{\rm d} & = & \left | \frac{V_{td}}{V_{ts}} \right | \; .
\end{eqnarray}
These simple results make the parametrization (21) uniquely favorable 
for the study of $B$-meson physics.

By use of current data on $|V_{ub}|$ and $|V_{cb}|$, i.e., $|V_{cb}| = 
0.039 \pm 0.002$ $^{10)}$ and $|V_{ub}/V_{cb}| =0.08 \pm 0.02$ $^{3)}$, we
obtain $\theta_{\rm u} = 4.57^{\circ} \pm
1.14^{\circ}$ and $\theta = 2.25^{\circ} \pm 0.12^{\circ}$. Taking
$|V_{td}| = (8.6 \pm 2.1) \times 10^{-3}$,
which was obtained from the analysis of current data on
$B^0_d$-$\bar{B}^0_d$ mixing,
we get $|V_{td}/V_{ts}| = 0.22 \pm 0.07$, i.e., $\theta_{\rm d} = 12.7^{\circ} 
\pm 3.8^{\circ}$.
Both the heavy quark mixing angle $\theta$ and the u-channel mixing
angle $\theta_{\rm u}$ are relatively small. The smallness of $\theta$ 
implies that Eqs. (24) and (25) are valid to a high degree of
precision (of order $1-c \approx 0.001$).
						
Recently a  global fit of these angles was made$^{17)}$, with rather
small uncertainties for the angles and the phase $\varphi $. One finds:
\begin{equation}
\begin{array}{cccccc}
\theta & = & (2.30 \pm 0.09)^{\circ} \; , \; & \theta_{\rm u} & = & (4.87 \pm 0.86)^{\circ}
\; , \nonumber \\
\theta_{\rm d} & = & (11.71 \pm 1.09)^{\circ} \; , \; & \varphi & = & (91.1 \pm 11.8 )^{\circ}
\; , \nonumber
\end{array}
\end{equation}
These values are consistent with the ones given above, however, the errors
are smaller.

f) According to Eq. (22), as well as Eq. (21), the phase $\varphi$ is
a phase difference between the contributions to $V_{us}$ (or $V_{cd}$) 
from the u-channel mixing and the d-channel mixing. Therefore
$\varphi$ is given by the relative phase of $D_{\rm d}$ and $D_{\rm
u}$ in the quark mass matrices (17), if the phases of $B_{\rm u}$ and
$B_{\rm d}$ are absent or negligible. 

The phase $\varphi$ is not likely to be $0^{\circ}$ or $180^{\circ}$, according
to the experimental values given above, even though the measurement of 
$CP$ violation in $K^0$-$\bar{K}^0$ mixing is not taken
into account. For $\varphi =0^{\circ}$, one
finds $\tan\theta_{\rm C} = 0.14 \pm 0.08$; and for $\varphi =
180^{\circ}$, one gets $\tan\theta_{\rm C} = 0.30 \pm 0.08$. Both
cases are barely consistent with the value of $\tan\theta_{\rm
C}$ obtained from experiments ($\tan\theta_{\rm C} \approx
|V_{us}/V_{ud}| \approx 0.226$). 

g) The $CP$-violating phase $\varphi$ in the flavor mixing matrix $V$ can be
determined from $|V_{us}|$ ($= 0.2205 \pm 0.0018$)
through the following formula, obtained easily from Eq. (21):
\begin{equation}
\varphi \; =\; \arccos \left ( \frac{s^2_{\rm u} c^2_{\rm d} c^2 +
c^2_{\rm u} s^2_{\rm d} - |V_{us}|^2}{2 s_{\rm u} c_{\rm u} s_{\rm d}
c_{\rm d} c} \right ) \; .
\end{equation}
\\
The two-fold ambiguity associated with the value of $\varphi$, coming
from $\cos\varphi = \cos (2\pi - \varphi)$, is removed if one
takes $\sin\varphi >0$ into account (this is required by current data on
$CP$ violation in $K^0$-$\bar{K}^0$ mixing (i.e., $\epsilon^{~}_K$). More
precise measurements of the angles $\theta_{\rm u}$ and
$\theta_{\rm d}$ in the forthcoming experiments of $B$ physics will
remarkably reduce the uncertainty of $\varphi$ to be determined from Eq.
(29). This approach is of course complementary to the direct determination of
$\varphi$ from $CP$ asymmetries in some weak $B$-meson decays into hadronic
$CP$ eigenstates. As mentioned above, the phase $\varphi $
appears to be very close to 90$^{\circ}$.

h) It is well-known that $CP$ violation in the flavor mixing matrix $V$ 
can be described in a way which is invariant with respect to phase changes 
by a universal quantity ${\cal J}$$^{18)}$:
\begin{equation}
{\rm Im} \left( V_{il} V_{jm} V^*_{im} V^*_{jl} \right) = {\cal J}
\sum\limits^{3}_{k,n=1} \left( \epsilon_{ijk}\epsilon_{lmn} \right ) \, .
\end{equation}
In the parametrisation (21), ${\cal J}$ reads
\begin{equation}
{\cal J} = s_{\rm u} c_{\rm u} s_{\rm d} c_{\rm d} s^2 c \sin \varphi \; .
\end{equation}
Obviously $\varphi = 90^{\circ }$ leads to the maximal value of ${\cal J}$.
Indeed $\varphi =90^{\circ}$, a particularly interesting case for $CP$ 
violation, is quite consistent with
current data. This possibility exists if $0.202 \leq \tan\theta_{\rm d} 
\leq 0.222$, or $11.4^{\circ} \leq \theta_{\rm d} \leq 12.5^{\circ}$.
In this case the mixing term
$D_{\rm d}$ in Eq. (17) can be taken to be real, and the term $D_{\rm 
u}$ to be imaginary, if ${\rm Im}(B_{\rm u}) = {\rm Im} (B_{\rm d})
=0$ is assumed. 
Since in our description of the flavor mixing the
complex phase $\varphi$ is related in a simple way to the phases of
the quark mass terms, the case $\varphi = 90^{\circ}$ is especially
interesting. It can hardly be an accident, and this case should be
studied further. The possibility that the phase $\varphi$ describing
$CP$ violation in the standard model is given by the algebraic number
$\pi/2$ should be taken seriously. It may provide a useful clue
towards a deeper understanding of the origin of $CP$ violation
and of the dynamical origin of the fermion masses.

The case $\varphi =90^{\circ}$ has been
denoted as ``maximal'' $CP$ violation$^{19)}$. It implies in our framework 
that in the complex
plane the u-channel and d-channel mixings are perpendicular to each
other. In this special case (as well as $\theta\rightarrow 0$), we have 
\begin{equation}
\tan^2\theta_{\rm C} \; =\; \frac{\tan^2\theta_{\rm u} ~ + ~
\tan^2\theta_{\rm d}}{1 ~ + ~ \tan^2\theta_{\rm u} \tan^2\theta_{\rm
d}} \; .
\end{equation}
To a good approximation (with the relative error $\sim 2\%$), 
one finds $s^2_{\rm C} \approx s^2_{\rm u} + s^2_{\rm d}$. 

\begin{figure}[t]
\hspace*{-1.8cm}\begin{picture}(400,160)(10,210)
\put(80,300){\line(1,0){150}}
\put(80,300.5){\line(1,0){150}}
\put(150,285.5){\makebox(0,0){$S_c$}}
\put(80,300){\line(1,3){21.5}}
\put(80,300.5){\line(1,3){21.5}}
\put(80,299.5){\line(1,3){21.5}}
\put(71,333){\makebox(0,0){$S_u$}}
\put(230,300){\line(-2,1){128}}
\put(230,300.5){\line(-2,1){128}}
\put(178,343.5){\makebox(0,0){$S_t$}}

\put(95,310){\makebox(0,0){$\gamma$}}
\put(188,309){\makebox(0,0){$\beta$}}
\put(109,350){\makebox(0,0){$\alpha$}}
\put(150,260){\makebox(0,0){(a)}}

\hspace*{-1.5cm}\put(300,300){\line(1,0){150}}
\put(300,300.5){\line(1,0){150}}
\put(370,285.5){\makebox(0,0){$s^{~}_{\rm C}$}}
\put(300,300){\line(1,3){21.5}}
\put(300,300.5){\line(1,3){21.5}}
\put(300,299.5){\line(1,3){21.5}}
\put(287,333){\makebox(0,0){$s_{\rm u} c_{\rm d}$}}
\put(450,300){\line(-2,1){128}}
\put(450,300.5){\line(-2,1){128}}
\put(395,343.5){\makebox(0,0){$s_{\rm d}$}}

\put(315,310){\makebox(0,0){$\gamma$}}
\put(408,309){\makebox(0,0){$\beta$}}
\put(329,350){\makebox(0,0){$\alpha$}}
\put(370,260){\makebox(0,0){(b)}}
\end{picture}
\vspace{-1.5cm}
\caption{The unitarity triangle (a) and its rescaled counterpart (b)
in the complex plane.}
\end{figure}
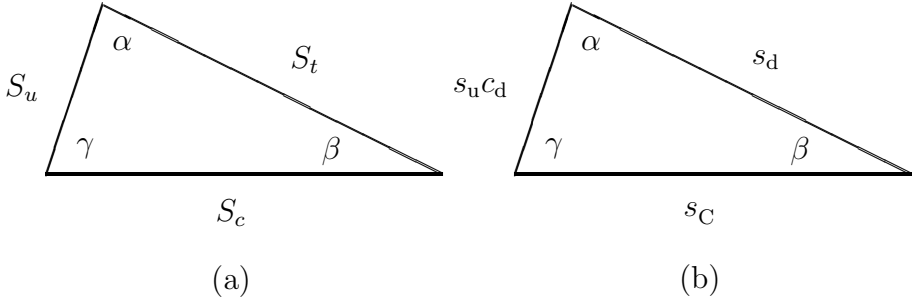
i) At future $B$-meson factories, the study of $CP$ violation will
concentrate on measurements of the unitarity triangle 
\begin{equation}
S_u ~ + ~ S_c ~ + ~ S_t \; = \; 0 \; ,
\end{equation}
where $S_i \equiv V_{id} V^*_{ib}$ in the complex
plane (see Fig. 2(a)). The inner angles of this triangle 
are denoted as usual:
\begin{eqnarray}
\alpha & \equiv & \arg (- S_t S^*_u ) \; , \nonumber \\
\beta  & \equiv & \arg (- S_c S^*_t ) \; , \nonumber \\
\gamma & \equiv & \arg (- S_u S^*_c ) \; .
\end{eqnarray}
In terms of the parameters
$\theta$, $\theta_{\rm u}$, $\theta_{\rm d}$ and $\varphi$, we obtain
\begin{eqnarray}
\sin (2\alpha) & = & \frac{2 c_{\rm u} c_{\rm d} \sin\varphi \left
( s_{\rm u} s_{\rm d} c + c_{\rm u} c_{\rm d} \cos\varphi \right )}{s^2_{\rm
u} s^2_{\rm d} c^2 + c^2_{\rm u} c^2_{\rm d} + 2 s_{\rm u} c_{\rm u} s_{\rm d} c_{\rm d} c
\cos\varphi} \; , \nonumber \\ \nonumber \\
\sin (2\beta) & = & \frac{2 s_{\rm u} c_{\rm d} \sin\varphi \left
( c_{\rm u} s_{\rm d} c - s_{\rm u} c_{\rm d} \cos\varphi \right )}{c^2_{\rm
u} s^2_{\rm d} c^2 + s^2_{\rm u} c^2_{\rm d} - 2 s_{\rm u} c_{\rm u} s_{\rm d} c_{\rm d} c
\cos\varphi} \; .
\end{eqnarray}
To an excellent degree of accuracy, one finds $\alpha \approx
\varphi$. In order to illustrate how accurate this relation is, let us
input the central values of $\theta$, $\theta_{\rm u}$ and $\theta_{\rm 
d}$ (i.e., $\theta = 2.25^{\circ}$, $\theta_{\rm u} = 4.57^{\circ}$
and $\theta_{\rm d} = 12.7^{\circ}$) to Eq. (35). Then one arrives at
$\varphi - \alpha \approx 1^{\circ}$ as well as $\sin (2\alpha)
\approx 0.34$ and $\sin (2\beta) \approx 0.65$. 
It is expected that $\sin (2\alpha)$ and $\sin (2\beta)$
will be directly measured from the $CP$ asymmetries in 
$B_d \rightarrow \pi^+\pi^-$ and $B_d \rightarrow J /\psi K_S$ modes
at a $B$-meson factory.

Note that the three sides of the unitarity triangle 
can be rescaled by $|V_{cb}|$. In a very good approximation
(with the relative error $\sim 2\%$), one arrives at
\begin{equation}
|S_u| ~ : ~ |S_c| ~ : ~ |S_t| \; \approx \; s_{\rm u} c_{\rm d} ~ : ~ 
s^{~}_{\rm C} ~ : ~ s_{\rm d} \; .
\end{equation}
Equivalently, one can obtain
\begin{equation}
s_{\alpha} ~ : ~ s^{~}_{\beta} ~ : ~ s_{\gamma} \; \approx \; s^{~}_{\rm C} 
~ : ~ s_{\rm u} c_{\rm d} ~ : ~ s_{\rm d} \; ,
\end{equation}
where $s_{\alpha} \equiv \sin\alpha$, etc.
The rescaled unitarity triangle is shown in Fig. 2(b). Comparing this
triangle with the LQ--triangle in Fig. 1, we find that they are 
indeed congruent with each other to a high degree of accuracy.
The congruent relation between these two triangles is particularly
interesting, since the LQ--triangle is essentially a feature of the physics
of the first two quark families, while the unitarity triangle is
linked to all three families. In this connection it is of special
interest to note that in models which specify the textures of the mass 
matrices the Cabibbo triangle and hence three inner angles of the unitarity
triangle can be fixed by the spectrum of the light quark masses and
the $CP$-violating phase $\varphi$$^{19)}$.

j) It is worth pointing out that the u-channel and d-channel mixing
angles are related to the Wolfenstein parameters$^{4)}$ in a simple way:
\begin{eqnarray}
\tan\theta_{\rm u} & = & \left | \frac{V_{ub}}{V_{cb}} \right | 
\; \approx \; \lambda \sqrt{\rho^2 + \eta^2} \; , \; \nonumber \\
\tan\theta_{\rm d} & = & \left | \frac{V_{td}}{V_{ts}} \right |
\; \approx \; \lambda \sqrt{ (1-\rho)^2 + \eta^2} \; ,
\end{eqnarray}
where $\lambda \approx s^{~}_{\rm C}$ measures the magnitude of $V_{us}$.
Note that the $CP$-violating parameter $\eta$ is linked to $\varphi$
through
\begin{equation}
\sin\varphi \; \approx \; \frac{\eta}{\sqrt{\rho^2 + \eta^2}
\sqrt{(1-\rho)^2 + \eta^2}} \; 
\end{equation}
in the lowest-order approximation. Then $\varphi =90^{\circ}$ implies
$\eta^2 \approx \rho ( 1- \rho)$, on the condition $0 < \rho < 1$. In
this interesting case, of course, the flavor mixing matrix can 
fully be described in terms of only three independent parameters.

k) Compared with the standard parametrization of the flavor mixing
matrix $V$ our parametrization has an additional
advantage: the renormalization-group evolution of $V$, from the weak
scale to an arbitrary high energy scale, is 
to a very good approximation associated only with the angle $\theta$. This
can easily be seen if one keeps the $t$ and $b$ Yukawa couplings only  
and neglects possible threshold effect in the one-loop
renormalization-group equations of the Yukawa matrices$^{20)}$.
Thus the parameters $\theta_{\rm u}$, $\theta_{\rm d}$ and $\varphi$
are essentially independent of the energy scale, while $\theta$ does
depend on it and will change if the underlying scale is shifted, say
from the weak scale ($\sim 10^2$ GeV) to the grand unified theory
scale (of order $ 10^{16}$ GeV). In short, the heavy quark mixing is
subject to renormalization-group effects; but the u- and d-channel
mixings are not, likewise the phase $\varphi$ describing $CP$
violation and the LQ--triangle as a whole.\\
\\
We have presented a new description of the flavor mixing 
phenomenon, which is based on the phenomenological fact that the quark 
mass spectrum exhibits a clear hierarchy pattern. This leads uniquely
to the interpretation of the flavor mixing in terms of a heavy quark
mixing, followed by the u-channel and d-channel mixings. The complex
phase $\varphi$, describing the relative orientation of the u-channel
mixing and the d-channel mixing in the complex plane, signifies
$CP$ violation, which is a phenomenon primarily linked to the physics
of the first two families. The Cabibbo angle is not a basic mixing
parameter, but given by a superposition of two terms involving the
complex phase $\varphi$. The experimental data suggest that the phase
$\varphi$, which is directly linked to the phases of the quark mass
terms, is close to $90^{\circ}$. This opens the possibility to
interpret $CP$ violation as a maximal effect, in a similar way as
parity violation.\\ 
\\
Our description of flavor mixing has many clear advantages compared
with other descriptions. We propose that it should be used in the
future description of flavor mixing and $CP$ violation, in particular, 
for the studies of quark mass matrices and $B$-meson physics.\\
\\
The description of the flavor mixing phenomenon given above is of special
interest if for the u- and d-channel mixings specific quark mass textures
are used.\\
In that case one often finds$^{14), 19)}$
apart from small corrections
\begin{eqnarray}
{\rm tan} \theta_{\rm d} & = & \sqrt{\frac{m_d}{m_s}} \; ,
\nonumber \\
{\rm tan} \theta_{\rm u} & = & \sqrt{\frac{m_u}{m_c}} \, .
\end{eqnarray}
The experimental value for ${\rm tan} \theta _u$ given by the ratio
$|V_{ub} / V_{cb}|$ is in agreement with the observed value for
$\left( m_u / m_c \right) ^{1/2} \approx 0.07$, but the errors for both
$\left( m_u / m_c \right) ^{1/2}$ and $|V_{ub} / V_{cb}|$ are the same
(about 25\%). Thus from the underlying texture no new information is
obtained.\\
\\
This is not true for the angle $\theta_{\rm d}$, whose experimental value
is due to a large uncertainty.: $\theta_{\rm d} = 12.7^{\circ } \pm 3.8^{\circ }$.
(The analysis given in$^{12)}$ indicates, however, that the uncertainty
for $\theta_{\rm d}$ may be less).
If $\theta_{\rm d}$ is given indeed by 
$(m_d /m_s)^{1/2}$, which is known to a high accuracy, 
we would know $\theta_{\rm d}$ and 
therefore all four parameters of the CKM matrix with high precision.\\
\\
The phase angle $\varphi $ is very close
to 90$^{\circ }$, implying that the LQ--triangle and the
unitarity triangle are essentially rectangular triangles$^{19)}$. In
particular the
angle $\beta $ which is likely to be measured soon in the study of the
reaction $B^0 \rightarrow J / \psi K^0_S$ is expected to be close
to $20 ^{\circ }$.\\
\\
Finally we should like to indicate the structure of the quark mass matrices
which follows if the two relations given above are fulfilled. Both for the
charge $2/3$ and charge $(- 1/3)$ quarks they have the structure:
\begin{equation}
M = \left( \begin{array}{lll}
	   0 & A & 0\\
	   A^* & B & C \\
	   0 & C^* & D
	   \end{array} \right)
\end{equation}
where $| A | \ll | B |, K| \ll | D | $. As discussed previously, the
texture zeros ate the $(1,3)$ and $(3,1)$--positions can always be
arranged by a suitable unitary transformation of the quark fields. Ones this
is carried out, a specific basis in flavor space is adopted. The relations $(40)$
follow if the $(1,1)$--element vanishes in this basis. It remains to be seen
what constitutes the deeper reason for this zero, perhaps a specific
reflection symmetry$^{14)}$ or simple irreducible term to
break the democratic symmetry among the quark flavors$^{21)}$.
$CP$--violation results if there is a relative phase between $A_d$ and $A_u$.
It is maximal if this phase is 90$^{\circ}$, as suggested by the
preliminary experimental data$^{22)}$.\\
\\
It will be very interesting to see whether the angles $\theta_{\rm d}$ and
$\theta_{\rm u}$ are indeed given by the square roots of the light quark mass
ration $m_d / m_s$ and $m_u /m_c$, which imply that the phase $\varphi$
is close to or exactly $90 ^{\circ }$. This would mean that the light quarks
play the most important r$\hat{\rm o}$le in the dynamics of flavor mixing and $CP$
violation and that a small window has been opened allowing the first view
accross the physics landscape beyond the mountain chain of the Standard
Model.\\
\\
Acknowledgements: It is my pleasure to thank Prof. A. Zichichi for the
invitation to come to Erice and for his continuous and fruitful efforts to
generate a lively and creative atmosphere at the Erice School on Subnuclear
Physics.\\
\\
\\
%

%
%
{\bf References}\\
\\
\hspace*{0.5cm} 1) Cabibbo,\, N. (1963): Phys. Rev. Lett. {\bf 10}

\, 2) Kobayashi,\, M., Maskawa,\, T. (1973): Prog. Theor. Phys. {\bf 49}, 652.

\, 3) Barnett,\, R.M. et al., Particle Data Group (1996):
Phys. Rev. {\bf D54}, 94.

\, 4) Wolfenstein,\, L. (1983): Phys. Rev. Lett. {\bf 51}, 1945

\, 5) Maiani,\, L. (1997): in {\it Proc. 1977 Int. Symp. on Lepton
and Photon Interactions\\
\hspace*{1cm} at High Energies} (DESY, Hamburg), 867
Chau,\, L.L., Keung,\, W.Y. (1984):\\
\hspace*{1cm} Phys. Rev. Lett. {\bf 53}, 1802
Fritzsch,\, H. (1985): Phys. Rev. {\bf D32}, 3058\\
\hspace*{1cm} Harari, H., Leurer,\, M. (1986): Phys. Lett. {\bf B181}, 123\\
\hspace*{1cm} Fritzsch,\, H.,  Plankl,\, J. (1987): Phys. Rev.
{\bf D35}, 1732

\, 6) Fritzsch,\, H., Xing,\, Z.Z. (1997): Phys. Lett. {\bf B413}, 396 

\, 7) Jarlskog,\, C. (1985): Phys. Rev. Lett. {\bf 55}, 1039

\, 8) Fritzsch,\, H. and Xing,\, Z.Z. (1989): Phys. Rev. {\bf D57}, 594

\, 9) Rasin,\, A.: Report No. hep-ph/9708216 (unpublished)

\hspace*{-0.3cm} 10) Fritzsch,\, H. (1987): Phys. Lett. {\bf B184}, 391

\hspace*{-0.3cm} 11) Fritzsch,\, H. (1987): Phys. Lett. {\bf B189}, 191

\hspace*{-0.3cm} 12) Hall,\, L.J.,  Weinberg,\, S. (1993): Phys. Rev. {\bf D48}, 979

\hspace*{-0.3cm} 13) Neubert,\, M. (1996): Int. J. Mod. Phys. {\bf A11}, 4173

\hspace*{-0.3cm} 14) Fritzsch,\, H. (1979): Nucl. Phys. {\bf B155}, 189

\hspace*{-0.3cm} 15) Dimopoulos,\, S., Hall,\, L.J., Raby,\, S. (1992):
Phys. Rev. Lett. {\bf 68}, 1984\\
\hspace*{-0.3cm} \hspace*{1.1cm} Barbieri,\, R., Hall,\, L.J.,\, Romanino,\, A. (1997):
Phys. Lett. {\bf B401}, 47

\hspace*{-0.3cm} 16) Fritzsch,\, H. (1977, 1978): Phys. Lett. {\bf B70}, 436; {\bf B73}, 317

\hspace*{-0.3cm} 17) Parodi,\, F., Roudeau,\, R., Stocchi,\, A.: Report No. hep--ph/9802289

\hspace*{-0.3cm} 18) Jarlskog,\, C. (1989): in {\it CP Violation}, edited by
C. Jarlskog (World Scientific), 3.

\hspace*{-0.3cm} 19) Fritzsch,\, H., Xing,\, Z.Z. (1995): Phys. Lett. {\bf B353}, 114

\hspace*{-0.3cm} 20) Babu,\, K.S., Shafi,\, Q. (1993):
Phys. Rev. {\bf D47}, 5004 (1993); and references therein

\hspace*{-0.3cm} 21) Lehmann,\, H., Newton,\, C., Wu,\, T.T. (1996): Phys. Lett. {\bf B384}, 249

\hspace*{-0.3cm} 22) Fritzsch,\, H., Xing,\, Z.Z.: in preparation.
\end{document}